\documentclass[aps,prd,showpacs,floatfix,
superscriptaddress,amsmath,preprintnumbers]{revtex4}

\usepackage{epsf}
\usepackage{amscd}              %
\usepackage{amsmath}            
\usepackage{bm}
\usepackage{latexsym}
\usepackage{graphicx}
\usepackage{comment}
\def\simge{
    \mathrel{\rlap{\raise 0.511ex
        \hbox{$>$}}{\lower 0.511ex \hbox{$\sim$}}}}
\def\simle{
    \mathrel{\rlap{\raise 0.511ex
        \hbox{$<$}}{\lower 0.511ex \hbox{$\sim$}}}}
\def\beqn{\begin{equation}}
\def\eeqn{\end{equation}}
\def\barr{\begin{eqnarray}}
\def\earr{\end{eqnarray}}
\def\bc{\begin{center}}
\def\ec{\end{center}}

\begin{document}
\preprint{DOE/ER/40762-373}
\title{Eikonal analysis of Coulomb distortion in quasi-elastic electron scattering}

\author{J. A. Tjon}
\affiliation{Physics Department, University of Utrecht, 3508 TA
Utrecht, The Netherlands }

\author{S.~J.~Wallace}
 \affiliation{Department of Physics, University of Maryland, College Park,
              MD 20742}

\pacs{24.10.-i, 
      25.30.Fj, 
      25.30.Hm 
}
\date{\today}

\begin{abstract}
 An eikonal expansion is used to provide systematic
corrections to the eikonal approximation through order $1/k^2$,
where $k$ is the wave number. Electron wave functions are obtained
for the Dirac equation with a Coulomb potential. They are used to
investigate distorted-wave matrix elements for quasi-elastic
electron scattering from a nucleus. A form of effective-momentum
approximation is obtained using trajectory-dependent eikonal phases and focusing
factors. Fixing the Coulomb distortion effects at the center of the nucleus,
the often-used ema approximation is recovered. Comparisons of these approximations are
made with full calculations using the electron eikonal wave
functions. The ema results are found to agree well with the full
calculations.
\end{abstract}

\pacs{24.10.-i,25.30.Fj,25.30.Hm}

\maketitle


\section{Introduction}
\label{sec:intro}

Professor Manoj Banerjee was a collaborator and friend of
both of the authors. It is an honor to contribute a paper that is dedicated to
his memory. He exhibited an enthusiasm and passion for physics that
has inspired many.

In this paper, we summarize some methods and results concerning the issue
of Coulomb corrections in quasi-elastic scattering of electrons by nuclei.
Electrons have been used extensively as an experimental probe of the internal
structure of nuclei in the past few
decades. Electron scattering is considered to be a precise tool in view of the
strength of the e.m. interaction, so that as a result the scattering
process can be treated in the one photon exchange approximation. In
particular, quasi-elastic scattering of electrons by nuclei has been used
to investigate properties like the validity of the
Coulomb sum rule in nuclei. Experiments have been performed at the
MIT Bates Laboratory ~\cite{Altemus83, Deady83, Hotta84, Deady86,
Blatchley86, Dytman86, Dow88, Yates93, Williamson97}, at the Saclay
Laboratory~\cite{Barreau83, Meziani84, Meziani85, Marchand85,
Zghiche94,Gueye99} and at SLAC~\cite{Baran88, Chen91, Meziani92}.
Although the electron may in general be considered a weak probe,
complications arise due to Coulomb distortion effects in the
electron wave function owing to the nuclear charge distribution. As
a result, in order to extract nuclear and nucleon structure
information from these experiments, the Coulomb distortion
contributions have to be accounted for in the theoretical analysis
of the data. In addition there exists the complication of the
presence of final state interaction. Studies have shown that in
the quasi-free region at high momentum transfer these effects are
expected to be small.

Neglecting the final state interaction, the Coulomb distortion can
be handled in the so-called distorted wave Born approximation. Here the
Coulomb potential is treated exactly by solving numerically the
Dirac equation in the presence of the Coulomb potential. Exact
solutions for the Dirac-Coulomb wave functions may be obtained as a
sum over partial waves.~\cite{Jin9294,Udias93,Udias95}. As the
electron energy increases, the partial-wave expansions converge more
slowly in spite of the fact that the Coulomb corrections become
smaller. These calculations are numerically  complex and have the
disadvantage of not allowing for a simple theoretical interpretation
in terms of nuclear structure functions. So one important
theoretical issue has been to investigate whether there is a
simple yet reliable way to characterize the reaction process in terms of
response functions, similar to what can be done in the
absence of final state interactions and Coulomb distortions.
Confining attention to high enough energy, a natural and
reliable framework for the description is given by the
use of the eikonal wave function.

The eikonal approximation clearly gets better at increasing energy
and it allows a simpler analysis of the effects of Coulomb
corrections. Some particulary transparent results have been obtained
using the eikonal approximation to derive an effective-momentum
approximation ($ema$)~\cite{Traini01,Traini88} that produces results
very similar to plane-wave results. It is important to include
focusing factors such as those found in the WKB
approximation~\cite{Yennie65}
and revisited more recently in quasi-elastic
scattering~\cite{Aste04a,Aste04b}. However, the attempts
to combine the eikonal analysis with focusing factors suffer
from the lack of a systematic basis. Significant
disagreements in the determination of nuclear response
functions from experimental data~\cite{Morgenstern01,Benhar06}
have arisen at least in part owing to the use of different
theoretical methods to remove the
Coulomb corrections. Therefore, it is of interest to study a
systematic expansion of the eikonal approach, where the various
effects arise in a natural way.

In order to address the issue of Coulomb
corrections, we have developed corrections to the eikonal
approximation based upon a systematic expansion in the high-energy
limit~\cite{TjW06}. This eikonal expansion is shown to be rapidly
convergent already at typical energies of few hundred MeV for
targets used in quasi-elastic scattering. Corrections to the eikonal
approximation have a long history. Work by Saxon and
Schiff~\cite{Saxon57} showed how to correct the approximation to
leading order in $1/k$. A systematic expansion for the scattering
t-matrix was developed by Sugar and Blankenbecler~\cite{Sugar69}.
Systematic corrections to the Glauber approximation~\cite{Glauber59}
were developed in~\cite{Baker72,Wallace73} and extended to the Dirac
scattering amplitude in~\cite{Wallace84}. However, a systematic
expansion for wave functions has not been developed prior to
Ref~\cite{TjW06}.

In Section~\ref{sec:eik}, we present the eikonal expansion for the
Dirac wave function and show that the focusing effect can be obtained at
order $1/k$ of the expansion. In particular, we focus on $u({\bf
r})$, which is a Pauli spinor containing the two upper components of
the Dirac wave function. The lower components are simply $\pm 1$ times
the upper components because of helicity conservation.
Convergence of the eikonal expansion is
shown to be fast in the few hundred MeV electron energy region.
Because there is a spin-orbit interaction, spin-dependent
terms arise in the eikonal expansion. They also are determined and
their effects are found to be negligibly small.
In Section~\ref{sec:QE} we summarize the basic formulae for
quasi-elastic electron scattering. In Section~\ref{sec:EMA} we deal
with effective-momentum approximations and discuss the original
$ema$ approach. A natural modification ($EMAr$) to the $ema$
approximation is proposed, where trajectory-dependent eikonal
phases and focusing factors are included.

Specializing to the longitudinal response,
section~\ref{sec:numerical} discusses quasi-elastic scattering by
use of a simple model of the nuclear response. Comparisons of the
full calculations of the response functions are made with the two
effective-momentum approximations. In the fits to the plane-wave
impulse approximation with an effective momentum small deviations
from unity of the normalization are found. However, overall
reasonable agreement is found with the full calculations, lending
support to the use of these effective-momentum approximations as a
basis for the theoretical analysis of the quasi-elastic data. Some
concluding remarks are made in Section~\ref{sec:remarks}.

\section{The Dirac electron wave function}
\label{sec:eik}

In this paper we consider electron scattering at intermediate
energies, where the eikonal approach is expected to be reasonably
accurate. The electron is assumed to be a Dirac particle. For the
Dirac equation, the eikonal expansion is carried out in two stages.
First we consider the Pauli spinor $u({\bf r})$ that contains the
two upper components of the Dirac wave function, i.e.,
\begin{equation}
\psi({\bf r}) = \begin{pmatrix} u({\bf r}) \cr \ell({\bf
r})\end{pmatrix}. \nonumber \label{eq:psi}
\end{equation}
It follows from the Dirac equation that the Pauli spinor $\ell({\bf
r})$ that contains the two lower components may be determined in a
second stage, where the two lower components are determined in terms
of $u({\bf r})$.

Eliminating the lower component spinor from the Dirac equation we
find for the upper-component spinor the equation
\begin{equation}
\Big( E_1 - V_c - \sigma \cdot {\bf p} \frac{1}{E_2 - V_c} \sigma
\cdot {\bf p} \Big)u({\bf r}), \label{eq:uofrDirac}
\end{equation}
where $E_1 = E - m$, $E_2 = E + m$ and E is the energy of the incoming particle.
For electron scattering it is generally the case that $E >> m$ and
thus $E_1 \approx E_2 \approx E$.  Because the electron mass is much smaller
than the energy, helicity is conserved and the lower components are
given simply by ${\bf \ell}_{\lambda}({\bf r}) = 2 \lambda u_{\lambda}({\bf r})$,
where $\lambda = \pm 1/2$ is the helicity.

For outgoing-wave boundary conditions, the Pauli spinor $u({\bf r})$
is written in terms of a complex eikonal phase
$\bar{\chi}^{(+)}=\chi^{(+)}({\bf r}) + i \omega^{(+)}({\bf r})$ and
a complex spin-dependent phase $\bar{\gamma}^{(+)} =
\gamma^{(+)}({\bf r}) + i \delta^{(+)}({\bf r})$ as follows
\begin{equation}
u^{(+)}({\bf r}) = \left( 1 - \frac{V_c}{E_2}\right) ^{1/2} e^{ik z
} e^{i \bar{\chi}^{(+)}}e^{i \sigma_e \bar{\gamma}^{(+)}}.
\label{eq:chi+}
\end{equation}
where $k=\sqrt{E^2-m^2}$ is the momentum of the incoming wave. The
wave propagates in the $z$-direction and an impact vector ${\bf b}$
is defined as the part of ${\bf r}$ that is perpendicular to the
$\hat{z}$-direction, i.e., ${\bf b} = \hat{z}\times ({\bf r} \times
\hat{z})$.  Three orthogonal unit vectors are : $\hat{z}$, $\hat{b}
= {\bf b}/|{\bf b}|$ and $\hat{e} = \hat{b} \times \hat{z}$. The
spin matrix in the eikonal phase is $\sigma_e = \sigma \cdot
\hat{e}$. The factor $(1 - V_c/E_2)^{1/2}$ is introduced in order to
sum up terms that otherwise arise in higher orders.

The eikonal expansion has been developed in Ref.~\cite{TjW06}. The
result is that the eikonal phases are expanded in a systematic
fashion in powers of $1/k$ as
\begin{eqnarray}
 \chi^{(+)} &=& \chi^{(+)}_0 + \chi^{(+)}_1 + \chi^{(+)}_2 + \cdots
\label{eq:chi_exp}
\nonumber \\
 \omega^{(+)} &=& ~~~~~~~~~~\omega^{(+)} _1 + \omega^{(+)} _2 + \cdots
\label{eq:omega_exp}
\nonumber \\
 \gamma^{(+)} &=& ~~~~~~~~~~\gamma^{(+)}_1 + \gamma^{(+)}_2 + \cdots
\label{eq:gamma_exp}
\nonumber \\
\delta^{(+)} &=& ~~~~~~~~~~~~~~~~~~~~\delta^{(+)}_2 +  \cdots ,
\end{eqnarray}
where the subscript of each term denotes the power of $1/k$ that is
involved. Explicit expressions can be found in Ref.~\cite{TjW06}.
The leading terms are given by
\begin{eqnarray}
\chi^{(+)}_0({\bf r}) &=& -\frac{1}{v} \int_{- \infty}^{z} dz'
V_c(r'),
\nonumber \\
\omega^{(+)}_1({\bf r}) &=& \frac{1}{2k} \int_{- \infty}^{z} dz'
{\nabla '}^2 \chi^{(+)}_0({\bf r}')
\nonumber \\
\gamma^{(+)}_1({\bf r}) &=& -\frac{1}{2k}  \int_{- \infty}^{z} dz'
\frac{\partial V_c(r)}{\partial b}
\nonumber \\
\delta^{(+)}_1({\bf r}) &=& 0, \label{eq:eik_out}
\end{eqnarray}
where higher order terms than $1/k$ have been dropped.

The upper-component spinor of the Dirac wave function for helicity
$\lambda$ and outgoing-wave boundary conditions is given by
\begin{equation}
u^{(+)}_{\lambda} =  f^{D(+)}_i({\bf r})\,e^{i k z} e^{i\chi^{(+)}}
e^{i \sigma_e \bar{\gamma}_i^{(+)}} \xi_{\lambda}, \label{eq:u+ofr}
\end{equation}
where $\xi_{\lambda}$ is a helicity eigenstate. The Dirac focusing
factor $f^{D(+)}$ is defined as
\begin{equation}
  f^{D(+)}_i({\bf r}) = \Biggr( 1 - \frac{V_c}{E_{2i}}\Biggr)^{1/2}
  e^{-\omega_i^{(+)}}.
  \label{eq:Dirac-focus}
  \end{equation}
One may work at various orders of the eikonal expansion by
truncating the expansions of Eq.~(\ref{eq:chi_exp}). Similarly, the
upper-component spinor for helicity $\lambda$ and incoming-wave
boundary conditions can be obtained by replacing in
Eqs.~(\ref{eq:eik_out},\ref{eq:Dirac-focus}) the superscripts $(+)$
by $(-)$ and the integration ranges $\int_{-\infty}^z$ by
$\int_z^{\infty}$.

Convergence of the eikonal expansion has been studied for scattering
of a 500 MeV electron. A rough estimate can be made of the higher
order corrections of the expansion. Given that the electron mass is
$m = .511$ MeV, it follows that $k \approx E$ and $v \approx 1$,
both within a part per million. The Coulomb potential is
approximately $V_c(0) = 25$ MeV at the center of the nucleus. The
eikonal expansion introduces corrections that involve the
nondimensional ratio $V_c/E \approx .05$, so we expect
$\frac{\chi^{(+)}_2}{\chi^{(+)}_0} \approx .0025$ It should be
noted, that the eikonal expansion is not convergent but is
asymptotic, meaning that the error should be bounded by the first
neglected term.

\begin{figure}[h]
\includegraphics[width=7cm,bb= 20 40 480 530,clip]{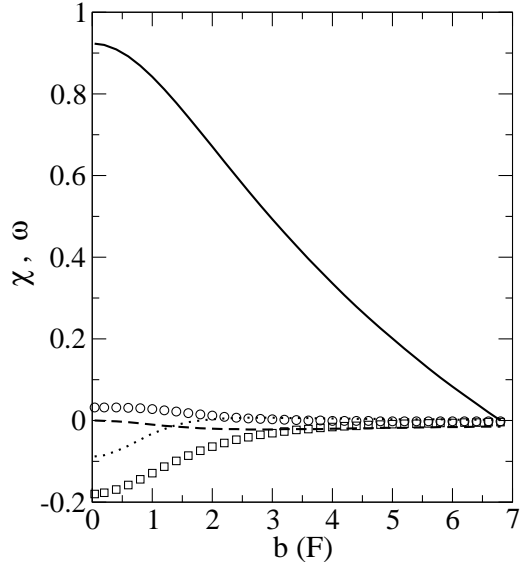}
\caption{Eikonal phases at z=0 versus impact parameter: solid line
shows $\chi_0$, dashed line shows $\chi_1$, dotted line shows
$\chi_2$, rectangles show $\omega_1$ and ovals show $\omega_2$. A
constant has been added to $\chi_0$ such that it vanishes at $b =
3.5R$.  Phases are shown for Z=100, E=200 MeV and R=2 fermi.
}\label{fig:EikPhases}
\end{figure}

Figure~\ref{fig:EikPhases} shows the eikonal phases for a charge
$Z=100$ and electron energy $E=200$ MeV. The Coulomb potential is
chosen to be
\begin{equation}
V_c(r) = - \frac{V_0 R}{\sqrt{r^2 + R^2}}, \label{eq:Vofr}
\end{equation}
where $-V_0$ is the value of the potential at $r=0$ and $R$ is a
range parameter. This Coulombic potential corresponds to a charge
density
\begin{equation}
\rho(r) = \frac{3V_0 }{4 \pi e} \frac{R^3}{(r^2 + R^2)^{5/2}}.
\end{equation}
The above
parameters were chosen in order to make the corrections visible. The
corrections are much smaller for a 500 MeV electron and a smaller
nuclear charge.

\section{Quasi-elastic electron scattering}
\label{sec:QE}

Let us consider the quasi-elastic nucleon knock-out process
$(e,e',N)$ from a nucleus. The cross section of quasi-elastic
electron scattering can be expressed in terms of the transition
matrix element ${\cal M}$:
\begin{eqnarray}
\frac{d \sigma}{d\Omega_f dE_f} = \int d\Omega_p \frac{4 \alpha^2
}{(2\pi)^5 }\overline{\left|{\cal M} \right|^2} E_f^2 p E_p,
\label{eq:dsigmadomega}
\end{eqnarray}
where $p$ is the momentum of the knocked-out nucleon and $E_p =
\sqrt{M^2 + p^2}$ is its energy. The bar denotes an average over
initial helicities and a sum over final helicities. We have
\begin{eqnarray}
{\cal M} = \delta_{\lambda_f\lambda_i} \int \frac{d^3q}{(2\pi)^2}
j_e^\mu \Biggr( \frac{1}{{\bf q}^2 - \omega^2}\Biggr) J^N_{\mu}({\bf
q},{\bf p}) \label{eq:Mtotal}
\end{eqnarray}
with $j_e^\mu$ the electron current matrix element
\begin{equation}
j_e^\mu =  \int d^3r 
u_{\lambda_f}^{(-)*}({\bf r})\gamma_{\mu}e^{-i {\bf q}\cdot{\bf r}}
u_{\lambda_i}^{(+)}({\bf r}), \label{eq:el_cur}
\end{equation}
for emission of a photon of energy $\omega = E_i - E_f$ and momentum
${\bf q}={\bf k_i -k_f}$. 
In Eq.~(\ref{eq:el_cur})
$u_{\lambda}^{(+/-)}$ are the electron wave functions 
 corresponding to initial momentum ${\bf
k}_i$ and final momentum ${\bf k}_f$ with
respectively outgoing and ingoing boundary conditions.
The helicity conservation factor $\delta_{\lambda_f \lambda_i}$
is produced by matrix elements that incorporate the
lower components.  Because the lower components are $\pm 1$ times
the upper components, in what follows one needs only the
upper components of the wave function.

In the plane-wave impulse approximation (PWIA), Coulomb distortion
of the electron waves is neglected and the electron is described by
a plane wave. As a result the integration over ${\bf r}$ produces
$\delta^{(3)}({\bf q} - {\bf Q})$ with ${\bf Q}= {\bf k_i -k_f}$. We
get
\begin{equation}
{\cal M}^{PWIA} = \delta_{\lambda_f \lambda_i} \frac{h_{PWIA}^{\mu}
J_{N \mu}({\bf Q},{\bf p})}{Q^2}, \label{eq:M_PWIA}
\end{equation}
where $h_{PWIA}^{\mu}$ denotes known helicity dependent
factors~\cite{TjW06}. The PWIA cross section may be expressed in
terms of longitudinal and transverse response functions, $R_L$ and
$R_T$
\begin{equation}
\frac{d \sigma}{d\Omega_f dE_f} = \sigma_{{\rm Mott}} \Biggr\{
\frac{Q^4}{{\bf Q}^4} R_L + \frac{Q^2}{2{\bf
Q}^2}\frac{1}{\epsilon}~ R_T\Biggr\},
\end{equation}
where
\begin{equation}
\sigma_{{\rm Mott}} = \frac{4\alpha^2 E_f^2 cos^2{\theta_e\over
2}}{Q^4},
\end{equation}
and
\begin{equation}
\epsilon = \Biggr[ 1 + \frac{2 {\bf Q}^2}{Q^2} tan^2 {\theta_e \over
2} \Biggr]^{-1}.
\end{equation}

With Coulomb corrections included, the longitudinal matrix element
of interest must take a gauge invariant form. This requires that the
electron current must be conserved in the sense that
\begin{eqnarray}
\int d^3r \Psi_{k_f}^{(-)*}({\bf r}) \Big(\omega j_e^0 - {\bf
q}\cdot {\bf j}_e\Big) e^{-i{\bf q}\cdot {\bf r}}
\Psi_{k_i}^{(+)}({\bf r}) = 0,
\end{eqnarray}
and that the nuclear current should separately be conserved,
\begin{eqnarray}
\int d^3r \Psi_{p}^{(-)*}({\bf r})e^{i{\bf q}\cdot {\bf r}} \Big(
\omega J_N^0 - {\bf q}\cdot {\bf J}_N\Big) \psi({\bf r}) = 0,
\label{eq:currcons}
\end{eqnarray}
where ${\bf q}$ is the photon three momentum.  With Coulomb
distorted waves, the photon momentum ${\bf q}$ differs from the
electron's momentum transfer ${\bf Q} = {\bf k}_i - {\bf k}_f$ and
the longitudinal current is defined with respect to the direction of
the photon that is exchanged, not with respect to the difference of
asymptotic electron momenta.

In view of current conservation, the longitudinal current matrix
element can be simplified to
\begin{eqnarray}
j_e^0 J_N^0 - (\hat{q}\cdot {\bf j}^e) (\hat{q}\cdot {\bf J}_N) =
j_e^0 J_N^0\Biggr( 1 - \frac{\omega^2}{{\bf q}^2}\biggr).
\label{eq:long_ME}
\end{eqnarray}

Using Eq.~(\ref{eq:u+ofr}), we get for the longitudinal
contribution to ${\cal M}$
\begin{eqnarray}
{\cal M}_L
 =&=& \delta_{\lambda_f\lambda_i}\int d^3r \int \frac{d^3q}{(2\pi)^2} e^{i ({\bf
Q}-{\bf q})\cdot {\bf r}} e^{i\chi({\bf r})} f^{D(-)}_f({\bf
r})f^{D(+)}_i({\bf r}) h_e ^{0}({\bf r})
 \Biggr( \frac{1}{{\bf q}^2}\Biggr)  J^N_{0}({\bf q},{\bf p}),
 \label{eq:Mlong}
\end{eqnarray}
where Eq.~(\ref{eq:long_ME}) has been used to include the components
of ${\bf j}_e$ and ${\bf J}_N$ that are parallel to ${\bf q}$.
The electron's momentum transfer is ${\bf Q} = {\bf k}_i - {\bf k}_f$ and $\chi =
\chi_f^{(-)}({\bf r}) + \chi_i^{(+)}({\bf r})$ includes the phases of
initial and final electron states. Note that
$\chi^{(+)}_i$, $\omega_i^{(+)}$, and $\bar{\gamma}_i^{(+)} =
\gamma_i^{(+)} + i \delta_i^{(+)}$ are obtained from
Eq.~(\ref{eq:eik_out}) with the z-axis parallel to initial momentum
${\bf k}_i$. In passing, we note that the Glauber approximation is
obtained when the eikonal phases for initial and final states are
evaluated using for each a z-axis parallel to the average momentum,
${1\over 2}({\bf k}_i + {\bf k}_f)$, and only the leading-order
phases, $\chi^{(+)}_0$ and $\chi^{(-)}_0$, are retained. This
approximation omits the focusing factors.

 The longitudinal response function is obtained by dividing the cross
section integrated over the angles of the knocked-out nucleon by the
Mott cross section,
\begin{equation}
R_L =\frac{{\bf Q}^4}{\sigma_{Mott}Q^4} \int d\Omega_p \frac{4
\alpha^2 }{(2\pi)^5 }\overline{\left|{\cal M}_L \right|^2} E_f^2 p
E_p, \label{eq:rlpwia}
\end{equation}
where ${\cal M}_L$ is the longitudinal amplitude of
Eq.~(\ref{eq:Mlong}). The full calculation thus involves a
six-dimensional integration in order to obtain the amplitude ${\cal
M}_L$.  Two more integrations over the angles of the knocked-out
nucleon are required in order to obtain the response function.
Results based on the eight-dimensional integration are called ``full
calculations'' in the following sections. The transverse transition
matrix element, which will not be considered further in this work,
is simply the difference of Eqs.~(\ref{eq:Mtotal}) and
(\ref{eq:Mlong}).

In the actual calculations we mostly use a very simple model for
the nuclear current
\begin{equation}
J_N^{\mu}({\bf q},{\bf p}) = \Biggr( \frac{p_i^{\mu} +
p_f^{\mu}}{\sqrt{4E_p(E_p-\omega)} }\Biggr) \psi({\bf q}-{\bf p}),
\label{eq:nuclearcurr}
\end{equation}
where $\hat{\psi}({\bf k})$ is a gaussian wave function for a bound
nucleon,
\begin{eqnarray}
\hat{\psi}({\bf k}) = (2 \pi \beta^2)^{3/4} e^{-\beta^2 k^2/4},
\label{eq:gaus}
\end{eqnarray}
normalized such that $\int d^3k |\psi({\bf k})|^2/(2\pi)^3 = 1$.

This simple gaussian model is used because the Coulomb corrections
should depend mainly on the electron wave functions. In order to get
some idea how a more realistic model of nuclear structure would
affect the results we considered also shell-model wave functions for
$^{56}Fe$ and $^{208}Pb$ nuclei. In those cases, the Coulomb
potential was calculated based on the empirical charge densities of
Ref.~\cite{de_Vries87} and the range parameter $R$ of the Coulomb
potential was determined so that the average Coulomb potential
matched the empirical one in the sense that $\int d^3
r\rho_{expt}(r) V_c(r) = \int d^3 r\rho_{expt}(r) V_{expt}(r)$. 
See Table~\ref{tab:HO&V0params} for the parameters used. For
the shell-model wave functions, the gaussian parameter $\beta$ was
selected such that the charge radius of the nuclei agreed with the
empirical charge radius and when the higher orbitals are included,
they are assumed to be described by harmonic-oscillator wave
functions in coordinate space as follows,
\begin{equation}
\psi_{nlm}({\bf r}) = N Y_{lm}(\Omega_r) r^l
\,_1F_1(-(n-l)/2,l+3/2,r\sqrt2/\beta)\, e^{-(r/\beta)^2},
\end{equation}
with normalization constants $N$ determined by  $\int d^3r
|\psi({\bf r})|^2 = 1$. Furthermore, $Y_{lm}$ are the well known
spherical harmonics and $_1F_1$ the confluent hypergeometric
functions. For the multi-shell studies we in addition use a more
realistic nuclear current of the form
\begin{equation}
j_N^{\mu} = K^{1/2} \bar{u}({\bf p}) \Big[ \gamma^{\mu} F_{1} +
\frac{i \kappa}{2M}F_{2} \sigma^{\mu \nu}q_{\nu} \Big] u({\bf p} -
{\bf q})\,\,\psi({\bf q}-{\bf p}) , \label{eq:fermionc}
\end{equation}
where $F_1(Q^2)$ and $F_2(Q^2)$ are nucleon form factors, $\kappa$
is the anomalous magnetic moment and $ K =  M^2/(E_pE_{{\bf p}-{\bf
q}})$ is a normalization factor arising from the spinors. For the
form factors $F_n$ standard dipole ones are taken.

The nuclear current Eq.~(\ref{eq:nuclearcurr}) is based upon the
current operator of a scalar nucleon with initial and final momenta
\begin{eqnarray}
p_f^{\mu} = \big( E_p, ~{\bf p} \big), \nonumber \\
p_i^{\mu} = \big( E_p - \omega, ~{\bf p}-{\bf q} \big),
\end{eqnarray}
where $\omega$ and ${\bf q}$ are the photon's energy and momentum.
Because of energy conservation, $E_p = M + \omega - B$, where $B
\approx .008 GeV$ is a typical binding energy of a nucleon. For the
PWIA response function the angular integration in
Eq.~(\ref{eq:rlpwia}) can easily be done. We find for the gaussian
model
\begin{eqnarray}
R_L^{PWIA}(\omega,{\bf Q})  =
 \frac{1}{\sqrt{2\pi}} \frac{
(2E_p-\omega)^2}{4(E_p-\omega)} \frac{\beta}{|{\bf Q}|} \biggr(
e^{-\beta^2( |{\bf Q}| - p)^2 /2} - e^{-\beta^2( |{\bf Q}| + p)^2
/2}\biggr).
\end{eqnarray}
$R_L^{PWIA}$ is normalized so that at fixed ${\bf Q}$, $ \int
d\omega R_L({\bf Q}, \omega)\approx 1 $.

\section{Effective-momentum approximations}
\label{sec:EMA}

Let us consider the electron current matrix element for emission of
a photon using spinors corresponding to
initial and final helicity $\lambda_i$ and $\lambda_f$. Using
Eq.~(\ref{eq:u+ofr}) and a similar relation for the incoming electron
wave function,  the electron current can be rewritten as
\begin{eqnarray}
&&j_e^{\mu} =  \int d^3ru_{\lambda_f}^{(-)*}({\bf
r})\gamma^{\mu}e^{-i
{\bf q}\cdot{\bf r}} u_{\lambda_i}^{(+)}({\bf r}) \nonumber \\
&&= \int d^3r \xi^{\dag}_{\lambda_f}
  e^{i({\bf Q} - {\bf q})\cdot {\bf r}} e^{ i \chi }
  e^{i\sigma_{e_f} \bar{\gamma}_f^{(-)*}} f^{D(-)}_f\gamma^{\mu}
f^{D(+)}_i e^{i \sigma_{e_i} \bar{\gamma}_i^{(+)}} \xi_{\lambda_i },
\label{eq:ufbarOui}
\end{eqnarray}
Using Eq.~(\ref{eq:ufbarOui}) the quasi-elastic transition matrix
element (\ref{eq:Mtotal}) takes the form
\begin{eqnarray}
{\cal M} = \delta_{\lambda_f\lambda_i} \int d^3r \int
\frac{d^3q}{(2\pi)^2} e^{i ({\bf
Q}-{\bf q})\cdot {\bf r}} e^{i\chi({\bf r})}
f^{D(-)}_f({\bf r})f^{D(+)}_i({\bf r}) h_e ^{\mu}({\bf r})
 \Biggr( \frac{1}{{\bf q}^2}\Biggr)  J^N_{\mu}({\bf q},{\bf p}),
 \label{eq:Mtotalf}
\end{eqnarray}
with
$$\delta_{\lambda_f,\lambda_i} h_e^{\mu} =
\xi^{\dag}_{\lambda_f} e^{i\sigma_{e_f} \bar{\gamma}_f^{(-)*}}
\gamma^{\mu} e^{i \sigma_{e_i} \bar{\gamma}_i^{(+)}}
\xi_{\lambda_i}.$$ We may now use a stationary-phase-like
argument to calculate Eq.~(\ref{eq:Mtotalf}). We see that for large
Q the integrand of (\ref{eq:Mtotalf}) has a rapidly changing phase
except when ${\bf Q - q + \nabla }\chi({\bf r}) = 0$. So we expect
that the dominant contribution in the integrals comes from around
this point.
Expanding the photon
propagator around ${\bf q} = {\bf Q}_{eff} \equiv {\bf Q} + {\bf
\nabla} \chi({\bf r}=0)$ we can explicitly factor the photon
propagator out of the integral
over ${\bf q}$. Hence we expect that Eq.~(\ref{eq:Mtotalf}) can
 in a good approximation for large $Q$ be determined by
\begin{eqnarray}
{\cal M}^{EMAr} = \delta_{\lambda_f\lambda_i} \Biggr( \frac{1}{{\bf
Q_{eff}}^2}\Biggr)   \int \frac{d^3r} {(2\pi)^2} e^{i {(\bf Q - q})\cdot {\bf r}}
 e^{i \chi({\bf r})} f^{D(-)}_f({\bf r})f^{D(+)}_i({\bf r}) h_e
^{\mu}({\bf r}) {\hat J}^N_{\mu}({\bf r},{\bf p}) \label{eq:EMAr}
\end{eqnarray}
with
$${\hat J}^N_{\mu}({\bf r},{\bf p}) = \int d^3q\,
e^{i {\bf q} \cdot {\bf r}}J^N_{\mu}({\bf q},{\bf p}). $$
Eq.~(\ref{eq:EMAr}) has obviously the form of an effective momentum
approximation except that the full ${\bf r}$-dependence of the eikonal
phase and the focusing factors is retained. It is
considerably simpler to calculate than the full six dimensional
integral for the quasi-elastic matrix element ${\cal M}$.

A further approximation can be made by
approximating the eikonal phase by $\chi \approx \chi(0) + {\bf
r}\cdot{\bf \nabla}\chi ({\bf 0})$ but keeping the r-dependence of
the focusing factors, as follows,
\begin{eqnarray}
{\cal M}^{EMAr'} = \delta_{\lambda_f\lambda_i} \Biggr( \frac{1}{{\bf
Q_{eff}}^2}\Biggr) e^{i \chi(0)} \int \frac{d^3r} {(2\pi)^2} e^{i {(\bf Q_{eff} - q})\cdot {\bf r}}
f^{D(-)}_f({\bf r})f^{D(+)}_i({\bf r}) h_e
^{\mu}({\bf r}) {\hat J}^N_{\mu}({\bf r},{\bf p}) \label{eq:EMAr'}
\end{eqnarray}
This is called the EMAr' approximation. Finally one may take
 both the eikonal phase and focusing factors
at the central value ${\bf r}=0$. In so doing we get the often-used $ema$
approximation. This approximation usually is
based on expanding the eikonal phase in a Taylor's series about
${\bf r} = {\bf 0}$ and keeping the first two terms. Moreover, the
focusing factors are approximated by their values at ${\bf r}= {\bf
0}$ and the helicity matrix elements are approximated by the
plane-wave values. Integration over ${\bf r}$ then gives
$\delta^{(3)}({\bf q} - {\bf Q}_{eff})$, so the longitudinal
amplitude simplifies to the PWIA form
\begin{equation}
{\cal M}_L^{ema} = 2\pi \delta_{\lambda_f \lambda_i} e^{i \chi(0)}h_{PWIA}^{0}
J_{N}^0({\bf Q}_{eff},{\bf p})\frac{f^{D(-)}_f({\bf
0})f^{D(+)}_i({\bf 0}) }{{\bf Q}_{eff}^2}. \label{eq:ML_ema}
\end{equation}

Combining the $e^{- \omega_i^{(+)}} \approx 1 - V_c(0)/(2E_i)$
factor of the eikonal correction with the $(1 - V_c/E_{2i})^{1/2}$
yields a focusing factor $f^{D(+)}_i \approx 1 - V_c/E_i$ in the
Dirac wave function, thus reproducing at $r=0$ the expected factor
$1 - V_c(0)/E_i$ that has been derived by Yennie, Boos and
Ravenhall~\cite{Yennie65} based on a WKB analysis of the
Dirac-Coulomb wave function. A similar result holds for the
final-state focusing factor, $f^{D(-)}_f$, which is approximately $1
-V_c/E_f$.  Thus, the overall focusing effect in the matrix element
is approximately equal to $(1 -V_c(0)/E_f)(1 -V(0)/E_i)$.

The effective momentum involves the gradient of the eikonal phase
shift $\chi = \chi_f^{(-)} + \chi_i^{(+)}$ at the origin. Because of
cylindrical symmetry of $\chi_i^{(+)}$ about the direction
$\hat{k}_i$, $\nabla \chi_i^{(+)}$ at the origin is nonzero only
along the direction $\hat{k}_i$, and similarly $\nabla \chi_f^{(-)}$
at the origin is nonzero only along the direction $\hat{k}_f$. With
$v_i = v_f \approx 1$, we find the same result as Traini,
\begin{equation}
{\bf Q}_{eff} = \hat{k}_i \Big[ k_i - \delta k \Big] - \hat{k}_f
\Big[ k_f - \delta k \Big], \label{eq:Qeff}
\end{equation}
where $\delta k = V_c(0)$. It is correct up to first order in the
eikonal expansion because the contribution from the gradient of
eikonal correction $\chi_1$ vanishes at the origin.

As shown by Rosenfelder~\cite{Rosenfelder80} and
Traini~\cite{Traini01}, there are significant cancellations in the
Coulomb corrections when response functions are evaluated in this
effective-momentum approximation ($ema$) using
the approximate focusing factors, $f^{D(+)}_i \approx 1 -
V_c(0)/E_i$ and $f_f^{D(-)} \approx 1 - V_c(0)/E_f$. Coulomb effects
in the focusing factors and the effective photon propagator cancel
if one considers the photon propagator of the transverse amplitude,
which is $1/[{\bf Q}_{eff}^2 - \omega^2]=1/[ 4 [k_i - V_c(0)][k_f -
V_c(0)]sin^2{1\over 2}\theta_e]$, i.e.,
\begin{equation}
\frac{ f^{D(-)}_f({\bf 0}) f^{D(+)}_i({\bf 0})}{{\bf Q}_{eff}^2 -
\omega^2} = \frac {1}{Q^2}. \label{eq:cancel}
\end{equation}
These factors produce the same result as in the
plane-wave case, Eq.~(\ref{eq:M_PWIA}),
but the momentum transfer argument in the nuclear structure
function is shifted.

Evidence has been presented that the momentum shift as predicted by
the $ema$ is too large and that a smaller value should be taken for
the Coulomb potential at the origin. This is done in view of the
plausible classical argument that the Coulomb potential which is felt by
the electron is not the central value of the potential,
but rather is the
average potential along the electron trajectory.
Based on this argument the momentum shift
$\delta k$ in the eikonal wave function is weakened by a factor
$f_{ema}$
\begin{equation}
\delta k = f_{ema} V_c(0),
\end{equation}
where $f_{ema}$ is determined by fitting the experimental
quasi-elastic peak value. In practice one finds a reduction factor
of typically $f_{ema}\approx 0.7$ to $0.8$.

It should be noted that although the $Q_{eff}$ is modified by the
factor $f_{ema}$, in the actual analysis one assumes that the
cancellation (\ref{eq:cancel}) still holds. In general, this
cancellation is clearly expected not to be complete. Using a
gauge-invariant response function as obtained from
Eq.~(\ref{eq:Mlong}) leads to deviations, which are of the order of
one percent. Another source of deviation is a more precise treatment
of the ${\bf r}$-dependence of the focussing factor as is done in
the EMAr approximation, given by Eq.~(\ref{eq:EMAr}). The break down
of the cancellation in the effective-momentum approach is reflected
in allowing for an additional overall normalization $A^{ema}$ in the
response function as given in Eq.~(\ref{eq:EMAfit}).

\section{Results}
\label{sec:numerical}
%
\begin{table}
\caption{Parameters used in calculations: $\beta$ is the harmonic
oscillator parameter; $V_0$ and $R$ are the Coulomb potential
parameters.  The gaussian model refers to Eq.~(\ref{eq:gaus}). }
\label{tab:HO&V0params}
\begin{tabular}{|cccc|}
\hline
 Nucleus  &  $~\beta$(fm)~ &~~$V_0$ (GeV) ~~&~~~$R$~(fm)~~~  \\
\hline
$\rm{gaussian}$~&~ 2.0  & 0.0273 &  2.0  \\
$^{208}Pb$~&~ 3.564  & 0.0256 &  7.10  \\
$^{56}Fe $ &  2.854  & 0.0124 &  3.97  \\
\hline
\end{tabular}
\end{table}
In this paper, we have described the eikonal expansion for
relativistic wave functions in the presence of a Coulomb potential
based on the Dirac equation. In the considered $1/k$ expansion,
focusing factors are obtained in a systematic manner by use of the
eikonal expansion. Although focusing factors take somewhat different
forms using the Klein-Gordon wave function, equivalent results are
found for the current matrix elements for the two
cases~\cite{TjW06}.

Calculations of the longitudinal response function are performed for
four cases: $PWIA$, $ema$ , EMAr and the full calculation using
distorted waves based on the Dirac equation. Eikonal phases are
evaluated through second order, i.e., $\chi = \chi_0 + \chi_1 +
\chi_2$ and $\omega = \omega_1 + \omega_2$. It should be noted
however, that the expansion converges rapidly for the parameters and
energies used and results based on $\chi_0 + \chi_1$ and $\omega_1$
differ by about 0.3\% at the quasi-elastic peak.

\begin{figure}[h]
\includegraphics[width=7cm,bb= 30 90 570
535,clip]{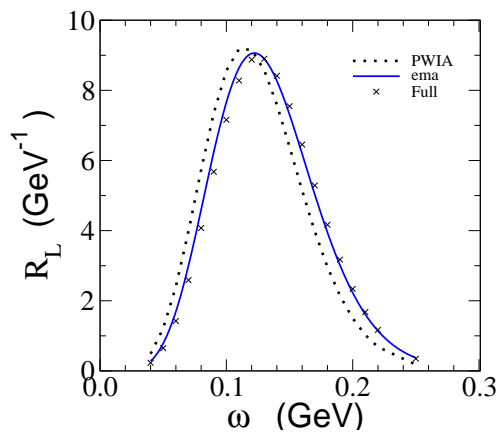}
\caption{Longitudinal response function
versus the electron's energy loss, $\omega$, calculated using the
gaussian model for $e^-$ scattering at $E= 500 MeV$ and $\theta_e
=60^o$. Dotted line shows $PWIA$, solid line shows $ema$ based on
$f_{ema}=0.7$ and x's show full calculations based on
Eq.~(\ref{eq:Mlong}).}\label{fig:RLn.e-.500}
\end{figure}
\begin{figure}[h]
\includegraphics[width=7cm,bb= 52  90 558 535,clip]{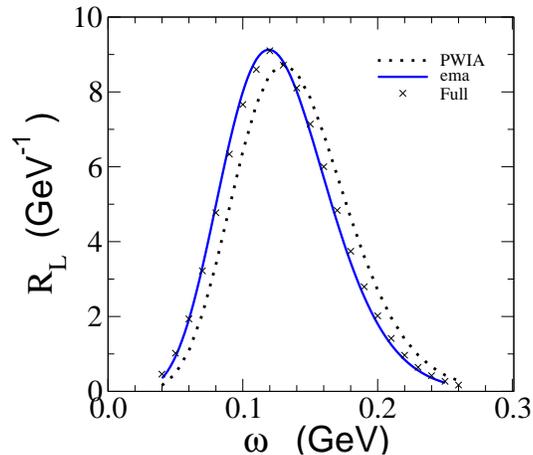}
\caption{Longitudinal response function versus the positron's energy
loss, $\omega$, calculated using the gaussian model for $e^+$
scattering at $E= 540 MeV$ and $\theta_e = 60^o$.  The dotted line
shows $PWIA$, solid line shows $ema$ using $f_{ema}=0.7$ and
$\times$'s show full calculations based on
Eq.~(\ref{eq:Mlong}).}\label{fig:RLn.e+.540}
\end{figure}

In the figures~\ref{fig:RLn.e-.500}-\ref{fig:RLn.e+.540} we have
used the gaussian model of Table~\ref{tab:HO&V0params} and have considered
electron and positron scattering. Calculations have omitted
final-state interactions of the knocked-out nucleon.
Figure~\ref{fig:RLn.e-.500} shows the longitudinal response function
for 500 MeV electrons and electron scattering angle $\theta_e=60^o$.
Here the full
calculations are plotted as x's, while the $ema$ calculation
based on $\delta k = 0.7\,\, V_c(0)$ is shown by the solid line. The
PWIA prediction is shown by the dotted line.
Figure~\ref{fig:RLn.e+.540} shows similar results for the
longitudinal response function
for $e^{+}$ scattering at 540 MeV using $f_{ema}=0.7$.

In general the $ema$ is seen in Figures~\ref{fig:RLn.e-.500} and
\ref{fig:RLn.e+.540} to produce a significant shift of $R_L$ away
from the PWIA result and towards the full calculation of $R_L$.
In both cases, the full calculations are
reproduced quite well by the $ema$ using $f_{ema}= 0.7$.
This reconfirms the findings of  Refs.~\cite{Morgenstern01,Aste04b}
that a smaller value of the $\delta k$ than $V_c(0)$ produces
better agreement with the full results. From the
effective-momentum approximation results we see that the shift in
momentum due to the Coulomb distortion is predicted to be opposite
in $e+$ to that of $e-$ scattering. There is indeed good agreement
between the response functions for $e-$ and $e+$ scattering at the
energies that make ${\bf Q}_{eff}$ close to the same for both.

As discussed in Ref.~\cite{TjW06}, it is possible to fit the
response functions more precisely if the momentum shift is allowed
to be a function of energy loss, $\omega$.  For the gaussian model,
it is found that near the quasi-free peak the momentum shift is well
described by $f_{ema} = 0.7$. The momentum shift can be
significantly larger in magnitude, corresponding to $f_{ema} > 1$,
when $\omega$ is significantly away from the value at the quasi-free
peak. The reason is that the Coulomb distortions tend to alter the
shape of the response functions away from the quasi-free peak.
However, simply using a constant $f_{ema}$ does not incur large
errors. The response integrated over $\omega$, as in the Coulomb sum
rule, is expected to be accurate within one or two percent.

In Table~\ref{table:EMAs} we show the numerical results for $e^-$
scattering at $E = 500 MeV$ and $\theta_e =60^0$  of the full
calculation together with the various effective-momentum
approximations, using the gaussian model. From this we see that
there is a close agreement between EMAr and the approximation
EMAr$'$, obtained from EMAr on replacing the eikonal phase by
$\chi(r) \approx f_{ema}~{\bf r}\cdot{\bf \nabla}\chi ({\bf 0})$.
This illustrates that the r-dependence of the phase shift can indeed
well be approximated by the linearized form. Moreover, both $ema$
and EMAr are in good agreement with the full result. It should be
noted that the assumed complete cancellation of the focus factor
(\ref{eq:cancel}) in the $ema$ approximation, which is found to hold
in this case, may be accidental.

\begin{figure}
\includegraphics[width=10cm,bb= 0 0 600 550,clip]{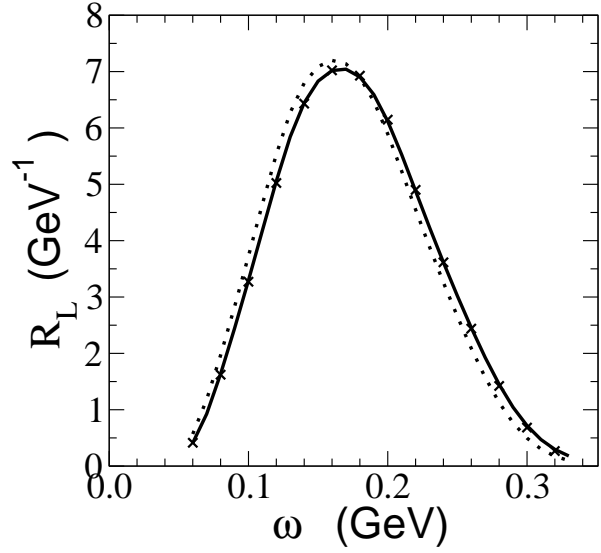}
\vspace {-1.5cm} \caption{ EMAr longitudinal response function for
$^{56}$Fe at E$_i$ = 500 Mev and $q=550$ MeV/c (solid line). The
corresponding PWIA response functions without Coulomb effects
included is shown by the dotted line.  A fit of the EMAr
response function using Eq.~(\ref{eq:EMAfit}) is shown by the
$\times$ symbols and values of the fitting parameters are given in
Table~\ref{table:EMAfits}. }\label{fig:RL.RUN1.Fe}
\end{figure}

\begin{figure}
\includegraphics[width=10cm,bb= 0 0 600 550,clip]{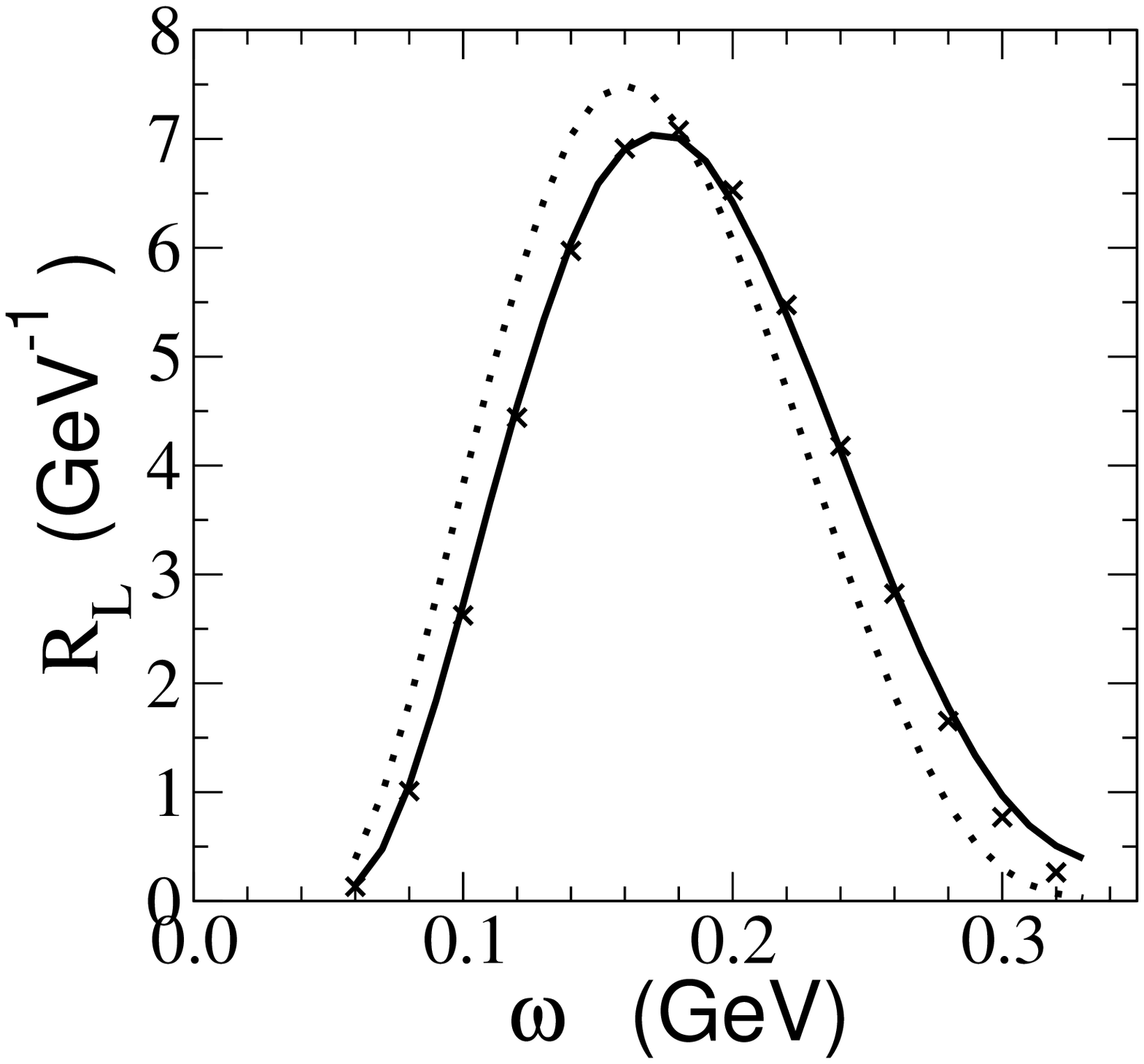}
\vspace {-1.5cm} \caption{ EMAr longitudinal response function for
$^{208}$Pb at E$_i$ = 500 Mev and $q=550$ MeV/c (solid line). The
corresponding PWIA response function without Coulomb effects
included is shown by the dotted line. A fit of the EMAr response
function using Eq.~(\ref{eq:EMAfit}) is shown by the $\times$
symbols and the parameters of the fit are given in
Table~\ref{table:EMAfits}. } \label{fig:RL.RUN1.Pb}
\end{figure}

The results for the simple gaussian model suggest that the $ema$ can
reproduce the results of the more elaborate EMAr analysis quite
well.  In order to test this proposition for a more realistic model
of the quasi-free scattering, calculations have also been made for
the $^{56}Fe$ and $^{208}Pb$ nuclei using shell-model wave functions
and a Coulomb potential that is based on the empirically determined
charge density. The Dirac nucleon current Eq.~(\ref{eq:fermionc}) is
used. Figures~\ref{fig:RL.RUN1.Fe} and \ref{fig:RL.RUN1.Pb} show the
results.

\begin{table}
\caption{The full calculation of the response function for the
gaussian model at various $\omega$ values for $e^-$ scattering at
500 MeV, $\theta_e = 60^o$, together with the effective-momentum
approximation results.}\label{table:EMAs}
\begin{tabular}{|ccccc|}
\hline
$~\omega$~~&~~Full~~&~~$ema$~~ &~~EMAr~~&~~EMAr$'$~  \\
\hline
 ~ 0.060  & 1.616 &  1.678 &  1.692 &  1.756 \\
 ~ 0.080  & 4.548 &  4.555 &  4.603 &  4.655 \\
 ~ 0.100  & 7.626 &  7.644 &  7.672 &  7.717 \\
 ~ 0.120  & 9.122 &  9.083 &  9.066 &  9.122 \\
 ~ 0.140  & 8.236 &  8.292 &  8.280 &  8.326 \\
 ~ 0.160  & 6.124 &  6.144 &  6.191 &  6.194 \\
 ~ 0.180  & 3.854 &  3.847 &  3.955 &  3.909 \\
 ~ 0.200  & 2.124 &  2.100 &  2.231 &  2.160 \\
 ~ 0.220  & 1.055 &  1.025 &  1.143 &  1.075 \\
\hline
\end{tabular}\label{table:EMAs}
\end{table}
\begin{table}
\caption{Parameters used to fit the EMAr response functions using
Eq.~(\ref{eq:EMAfit}).}\label{table:EMAfits}
\begin{tabular}{|cccccc|}
\hline
Nucleus     &~~$E$~~&~~$q$~~ &~~$\delta$k~~&~~$f_{ema}$~~&~$A$~  \\
\hline
$^{58}Fe$~~ &~~0.5~~&~~0.55~~&  -8.8 &  0.71  &  0.99   \\
$^{208}Pb$~~&~~0.5~~&~~0.55~~&  -21.0 &  0.82  &  0.98    \\
\hline
\end{tabular}\label{table:EMAfits}
\end{table}
%

\begin{figure}
\includegraphics[width=10cm,bb= 0 0 600 550,clip]{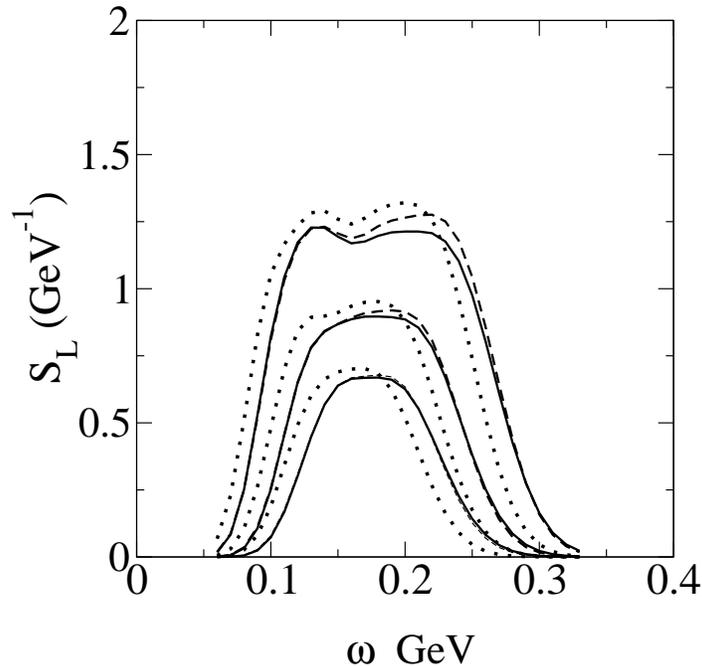}
\vspace{-0.2cm} \caption{ Solid lines show the partial EMAr
longitudinal response function for the $1p$, $1d$ and $1g$ shells of
$^{208}$Pb at E$_i$ = 500 Mev and $q=550$ MeV/c. The corresponding
$ema$ response functions based on $f_{ema} = 0.82$ are shown as dash
lines
 and the PWIA response functions are shown as dotted lines.
The upper group is the $1g$ shell, the middle group is the $1d$
shell and the lower group is the $1p$ shell. }
\label{fig:RL.Pb.1p-1g.EMAr.EMA.PW}
\end{figure}
In these figures, we show the EMAr calculations as solid lines and
two-parameter fits to them using the $ema$-fitting formula of
Eq.~(\ref{eq:EMAfit}) as $\times$'s.
\begin{equation}
R_L({\bf Q},\omega)= A^{ema}~ R_L^{PWIA}({\bf Q}_{eff},\omega)
\label{eq:EMAfit}
\end{equation}
The obtained parameters are given in Table~\ref{table:EMAfits}. The
main fitting parameter is the value of $f_{ema}$ that is used to
determine the momentum shift in ${\bf Q}_{eff}$.  Using the
shell-model wave functions, we find that response functions are fit
by $f_{ema}=0.71$ for $^{56}Fe$ and by $f_{ema}=0.82$ for
$^{208}Pb$.  Because of small distortions of the shape of the EMAr
response function relative to the shape of the PWIA response
function, a minor change of normalization is used also, as given by
the $A$ parameter. The Coulomb effects are larger for the $^{208}Pb$
nucleus because of the larger Coulomb potential, however the results
can be fit using the $ema$ formula with an appropriate value of
$f_{ema}$.  The fact that similar values of $f_{ema}$ are found for
the gaussian model, for $^{56}Fe$ and for $^{208}Pb$ demonstrates
that the Coulomb corrections are not very sensitive to the nuclear
model.

Figure~\ref{fig:RL.Pb.1p-1g.EMAr.EMA.PW} shows partial response
functions for individual shells of $^{208}Pb$ at 500 MeV electron
energy. Three shells, the $1p$, $1d$ and $1g$ shells of $^{208}Pb$,
are shown based on three calculations: the EMAr, $ema$ using
$f_{ema}=0.82$ and PWIA.  The results for individual shells show
that the $ema$ results are close to the EMAr response functions that
take into account the r-dependence of eikonal and shell wave
functions. Differences are somewhat larger than for response
functions summed over all shells. This is expected because the
radial wave functions for the shells with $\ell >0$ are suppressed
near the origin where the Coulomb potential is largest.  The $ema$
uses an average value for the Coulomb potential that works well for
the sum over all shells and is less accurate for individual shells.

\section{Concluding remarks}
\label{sec:remarks}

Within the eikonal approach we have studied how well the Coulomb distortion
effects are described by effective-momentum approximations
in electron scattering on nuclei at intermediate energies.
This is expected to be a reliable description at increasing high
energy. We have shown using a systematic eikonal expansion that
already in the few hundred $MeV$ region the convergence of the
expansion is indeed very fast and that the leading orders of the
eikonal phase and focus factors in the electron wave function are
sufficient to describe the Coulomb distortion in an accurate way.
Moreover, the effective-momentum approximations are found to agree
well with the full eikonal based calculations.

In this paper we have focused on the longitudinal response function.
The transverse response contribution has also been calculated
together with the spin-dependent terms occurring in the eikonal wave
function~\cite{TjW08}. From the present study we find strong support
for the conjecture that the effective-momentum approach can be used
as the basis for analysis of the inclusive experimental data. In
particular, the $ema$ approximation as used in the actual analysis
of the experimental data is found in our model studies to do well.

A more precise form of the effective-momentum approximation would be
useful for removing Coulomb corrections from experimental data in a
straightforward manner and has been suggested in Ref.~\cite{TjW06}.
In order to have a precise result, one could determine appropriate
values of the momentum-shift function $\delta
k(k_i,\omega,\theta_e)$ from which the appropriate ${\bf Q}_{eff}$
may be calculated as in Eq.~(\ref{eq:Qeff}).

In the $ema$ approximation a weakening factor $f_{ema}$ for the
momentum shift is usually introduced, which takes into account in a
phenomenological way the trajectory dependence of the phase shift.
We have studied in this paper the EMAr approximation, which includes
explicitly this r-dependence in the eikonal shift and focus factor
without the introduction of $f_{ema}$. Up to a possibly small
overall normalization constant correction, this is found to be in
good agreement with the full calculation

Our results for response functions omit final state interactions of
the knocked-out nucleon. Their inclusion would affect the shape of
the response functions but not the value of the Coulomb sum rule,
which involves a sum over a complete set of states. The Coulomb
corrections found in this work are not large enough to explain the
differences that have been reported for the Coulomb sum rule by
different experimental groups.\cite{Williamson97, Meziani84,
Meziani85}

\acknowledgements
 This work was supported by the
U.S. Dept. of Energy under contract 
DE-FG02-93ER-40762.

\bibliography{basename of .bib file}

\end{document}